\documentclass[11pt]{iopart}

\usepackage{iopams}
\usepackage{cite}
\usepackage{graphicx}
\usepackage{endfloat}
\begin{document}
\title{Carrier mean free path and temperature imbalance in mesoscopic wires}
\author{M. A. Kuroda$^1$, J.-P. Leburton$^{1,2}$}
\address{$^1$ Beckman Institute and Department of Physics, University of
Illinois at Urbana-Champaign, IL 61801, USA.}
\address{$^2$ Department of Electrical and Computer Engineering, University of Illinois at Urbana-Champaign, IL 61801, USA.}
\eads{\mailto{mkuroda@illinois.edu}, \mailto{jleburto@illinois.edu}}

\begin{abstract}
Non-coherent electronic transport in metallic nanowires exhibits different carrier temperatures for the non-equilibrium forward and backward populations in the presence of electric fields. Depending on the mean free path that characterizes inter-branch carrier backscattering transport regimes vary between the ballistic and diffusive limits. In particular, we show that the simultaneous measurements of the electrical characteristics and the carrier distribution function offer a direct way to extract the carrier mean free path even when it is comparable to the conductor length. Our model is in good agreement with the experimental work on copper nanowires by Pothier {\it et al.} [Phys. Rev. Lett. {\bf 79}, 3490 (1997)] and provides an elegant interpretation of the inhomogeneous thermal broadening observed in the local carrier distribution function as well as its scaling with external bias.
\end{abstract}

\pacs{73.23.-b, 65.80.+n, 73.21.Hb, 73.63.Fg, 73.63.Nm}
\vspace{2pc}
\noindent{\it Keywords}: diffusive transport, nanowires, nanotubes, mean free path

\maketitle

Present understanding of transport in mesoscopic systems relies on two different approaches: Landau's theory of Fermi liquids \cite{baymlandau} and Tomonaga-Luttinger liquids (TLL) theory \cite{solyom1979}. The latter describes correlated one dimensional (1D) systems and is characterized by a power law decay at the Fermi level at $T=0$, instead of the discontinuity observed in Fermi liquids. Recent progresses in fabrication technology have made available a variety of material structures to study the electronic properties in quasi-1D systems like quantum wires \cite{hu1999}, nanotubes \cite{iijima1991} and nanoribbons \cite{tapaszto2008}. Despite renewed experimental efforts, TLL features of 1D systems have yet to be indisputably proven, as their manifestation is limited fundamentally by their sensitivity to disorder and surface roughness, and the fact that inherent perturbations induced by the measurement should be much smaller than thermal fluctuations. These undesired effects cause loss of coherence amongst particles, and transport becomes diffusive.

In the past, transport experiments on copper nanowires at low temperature \cite{pothier1997}
 have shown to exhibit quasi-particle distributions with a two-step profile, a shape expected in a regime with no carrier interactions due to the superposition of the distribution functions in the leads \cite{nagaev1995}. However, the thermal broadening of the local distribution function (hot carrier effects) suggested significant carrier scattering, thereby invalidating lack of interactions.

In this paper we show that in non-coherent transport and beyond the diffusive limit the non-equilibrium carrier distribution in 1D-systems cannot be described by a single energy distribution but as a superposition of two distinct (forward and backward) carrier populations coupled by mutual scattering. Depending on the strength of the coupling between the two populations, the transport regimes varies between the well-known ballistic and diffusive limits. The model provides a straight correlation between the non-uniform thermal broadening of the carrier distribution and the mean-free path (even when the latter is comparable to the channel length) and presents good agreement with the experimental work in mesoscopic copper wires \cite{pothier1997}, describing both the inhomogeneous local thermal broadening as well as the scaling law of the carrier distribution function observed the high-bias. We also discuss the conditions for the observation of this phenomena in other material systems.

\begin{figure}[htpb]
\hspace{1in}
\includegraphics[width=4in]{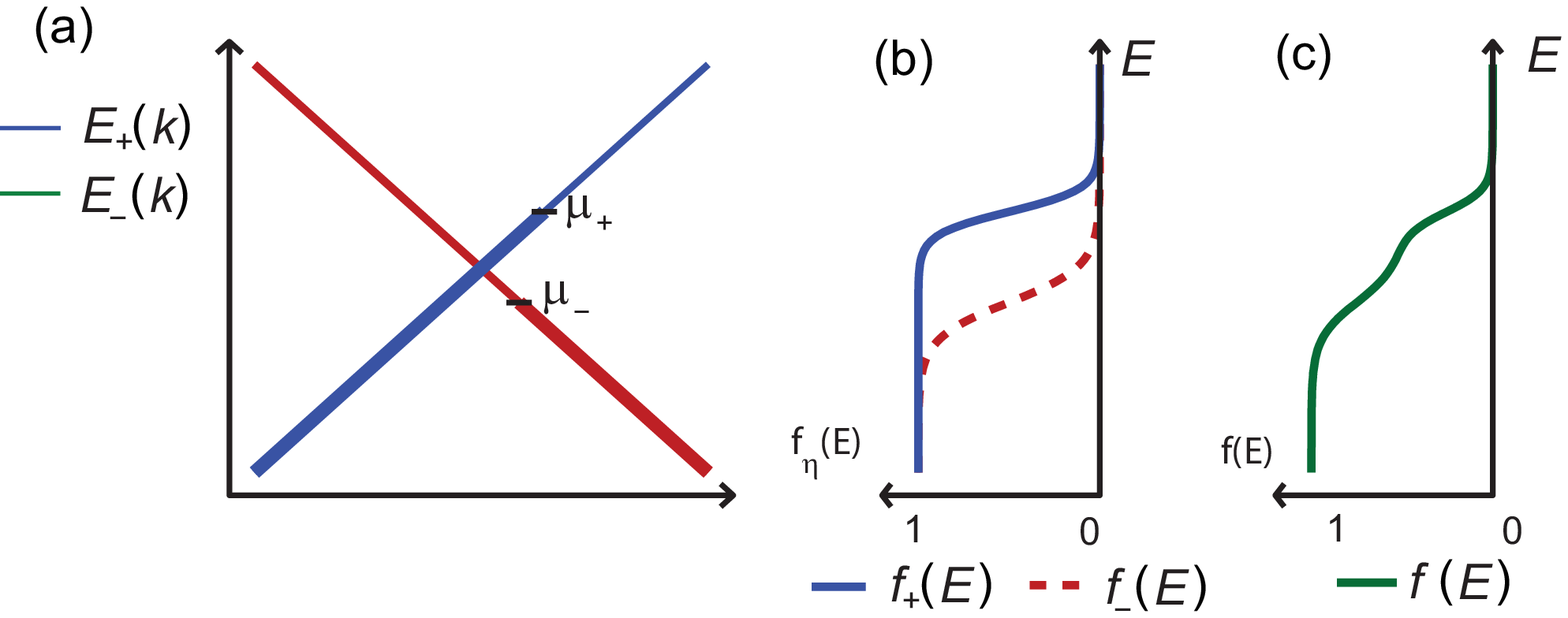}
\caption{(a) Band structure of two branch system level. (b) Forward and backward carrier distribution function out of thermal equilibrium in the presence of a field.
(c) Effective quasi-particle distribution function.} \label{fig:distrib}
\end{figure}

We assume that close to the Fermi level the energy dispersion in the 1D conductor is well described by linear branches $E_\pm(k)$\footnote{This is a good approximation in metallic conductors regardless the number of bands as far as no bottom of any subconducting band lies close the Fermi level, while it is the realistic band structure of metallic carbon nanotubes}, as shown in Fig.~\ref{fig:distrib}.a. For simplicity we only consider two branches, but our model and conclusions can be extended to mesoscopic systems with multiple bands provided that, as we show later, in 1D conductors current and heat flow do not depend on the branch Fermi velocity of the system. We assume that each of these two branches exhibits a $2g_c$-degeneracy (where the factor of 2 accounts for the spin). We group the carrier populations according to the sign of the Fermi velocity ($v_F= \hbar^{-1} \partial_k  E_\pm$). The effective intra-branch electron-electron (e-e) scattering thermalizes the distribution (i.e.~$\tau_{e-e}^{intra}\rightarrow 0$) causing the loss of coherence . Hence, each of these populations is described by a Fermi distribution function $f_\eta(E)$ (with $\eta = +,-$) \cite{kuroda2008}. In the presence of an electric field a population imbalance between the branches arises because of inefficient inter-branch carrier scattering, which creates a quasi-Fermi level difference ($\mu_+ \neq \mu_-$) and disrupts the thermal equilibrium between the two populations ($T_+\neq T_-$) as depicted in Fig.~\ref{fig:distrib}.b. Under these conditions we have recently shown that the net carrier and heat transport is expressed as \cite{kuroda2008}:
\begin{eqnarray}
I = g_c G_q \frac{\mu_+-\mu_-}{e}\label{eq:current}\\
U = \frac{g_c}{2} \left(G_{th}^+ T_+ -G_{th}^- T_-\right)\label{eq:heatflow}
\end{eqnarray}
in terms of the quantum electric ($G_q = e^2/(\pi \hbar) $) \cite{datta} and thermal ($G_{th}^\pm = \pi k_B^2 T_\pm /(3\hbar)$) \cite{rego1997} conductance, respectively. Neither the current nor the heat flow depend on the magnitude of the branch Fermi velocity because of the system dimensionality. Because of the constant density of states, the local carrier distribution function measured experimentally \cite{pothier1997,pothier1997b} is the average of the branch distribution functions :
\begin{equation}
f(E) = \frac{1}{2}\left[f_+(E)+f_-(E)\right] \label{eq:distfunc}
\end{equation}
as shown in Fig.~\ref{fig:distrib}.c. Two steps in he distribution function are clearly observed when $|\mu_+-\mu_-| \gg k_BT_+,k_BT_-$.

\begin{figure}[htpb]
\hspace{1in}
\includegraphics[width=4.5in]{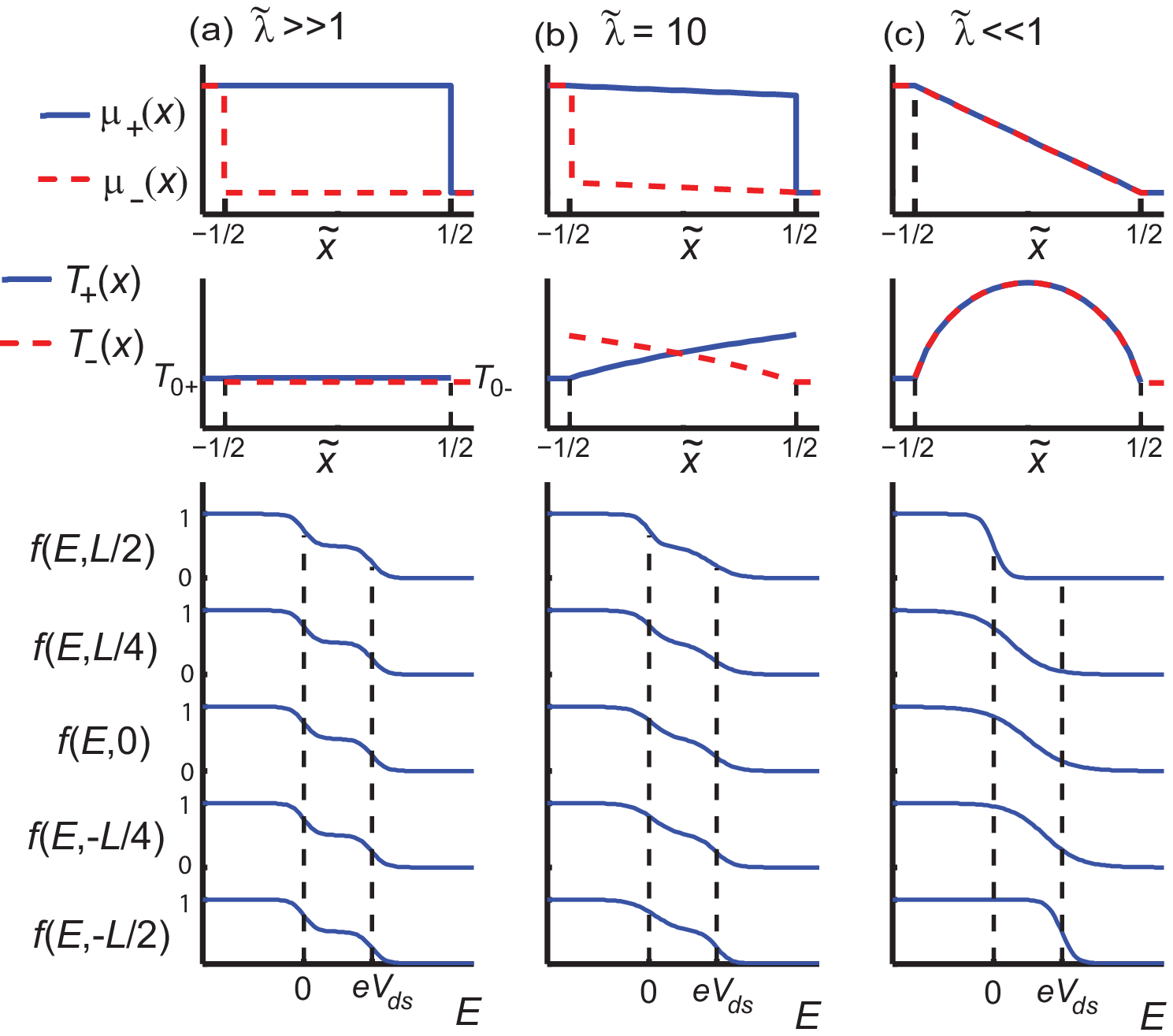}
\caption{Quasi-Fermi level (top) and temperature (middle) profiles for forward
and backward populations in different transport regimes: (a)
Ballistic $\tilde{\lambda} \gg 1$ (b) $\tilde{\lambda} = 10$ and
(c) Diffusive $\tilde{\lambda} \ll 1$. Bottom: quasi-particle
distribution function at $\tilde{x}$ = -1/2, -1/4, 0, 1/4 and 1/2 for the same regimes as above.} \label{fig:schemes} 
\end{figure}

We denote $\lambda$ the mean free path characterizing interactions amongst carriers in {\it different} branches, which tends to restore the equilibrium between the two populations. We assume these interactions involve inter-branch e-e, impurity or acoustic phonon (if the sound velocity $v_s\ll v_F$) processes, and only induce quasi-particle backscattering, i.e. no energy is transferred from the carrier populations to the external system. In this case the effective electric field $F$ along the channel has been shown to be \cite{kuroda2008}:
\begin{equation}
F = \frac{I}{g_c G_q \lambda} \label{eq:field}.
\end{equation}
By integrating this equation along the channel, assuming that the mean free path remains constant and using Eq.~\ref{eq:current}, we find the drain-source bias voltage $V_{ds}$:
\begin{equation}
V_{ds} = V_{c}+V_{ch} = \frac{\mu_+-\mu_-}{e}
\left(1+\frac{L}{\lambda}\right) \label{eq:Vds}.
\end{equation}
The magnitudes $V_c$ and $V_{ch}$ denote voltage drops at the contacts and along channel, respectively, where the former is due to the quantum contact resistance \cite{datta}. We have also shown that the temperature profiles for forward and backward populations are given by:
\begin{equation}
\pm G_{th}^\pm \partial_xT_\pm = \frac{I F}{2} \mp
\frac{U}{\lambda}
\end{equation}
under the influence of quasi-particle backscattering. In particular, direct integration of this equation with the boundary conditions $T_+(-L/2) = T_{0+}$ and $T_-(L/2)
= T_{0-}$ (perfectly absorbing contacts) yields:
\begin{eqnarray}
T_\pm(x) = \sqrt{T_{0\pm}^2\pm\frac{(1/2\pm \tilde{x}) \left(T_{0-}^2 -
T_{0+}^2\right)}{(1 + \tilde{\lambda})}+ \frac{\left(1/2 \pm
\tilde{x}\right) (1/2 \mp \tilde{x} +
\tilde{\lambda})}{(1+\tilde{\lambda})^2 } \frac{V_{ds}^2
}{\mathcal{L}}}\label{eq:tempprof}
\end{eqnarray}
where $\mathcal{L}=\frac{\pi^2}{3} \left(\frac{k_B}{e}\right)^2$ is the Lorenz number. The variables $\tilde{\lambda}$ and $\tilde{x}$ stand for the rescaled position ($\tilde{x} = x/L$) and mean free path  ($\tilde{\lambda}=\lambda/L$), respectively. The strength of the interaction sets two limiting regimes: (i)  ballistic ($\tilde{\lambda} \gg 1$) and
(ii) diffusive ($\tilde{\lambda}\ll 1$). In Fig.~\ref{fig:schemes} we compare the quasi-Fermi levels (top), temperatures profiles (middle) and local energy distribution function different positions along the channel (bottom) for these two limits and the intermediate case with $\tilde{\lambda} = 10$. For illustrative purposes, the bias voltage $V_{ds}=100\times k_BT_{0-}$ and the asymmetric boundary conditions $k_BT_{0+} = 1.1 \times k_BT_{0-}$ are used. On
the one hand, in the case of ballistic transport the quasi-Fermi levels for forward and backward populations remain constant along  the conductor (Fig.~\ref{fig:schemes}.a). As a result, the electric field vanishes along the channel and the drain-source voltage $V_{ds}$ is reduced to the voltage drop at the contacts:
\begin{equation}
V_{ds} = \frac{(\mu_+-\mu_-)}{e}  \label{eq:vdsball}
\end{equation}
In addition, the distribution of forward (backward) carriers exhibits a homogeneous temperature profile along the channel with the value of the left (right) lead temperature. In this regime, the local (average) distribution function $f(E)$ (Eq.~\ref{eq:distfunc}) exhibits two pronounced steps independently of the position in the wire when $eV_{ds} \gg k_BT_{0\pm}$. On the other hand, if $\tilde{\lambda}\ll 1$  (Fig.~\ref{fig:schemes}.c) a single Fermi-Dirac distribution describes the energy quasi-particle distribution since the quasi-Fermi level difference and thermal imbalance between carrier populations become negligible due to the effective inter-branch backscattering. Indeed, the drain-source voltage is due to the voltage drop along the channel (directly proportional to the channel length) as $V_c$ can be neglected (Eq.~\ref{eq:Vds}). In this regime, the thermal broadening along the channel (Eq.~\ref{eq:tempprof}) reads:
\begin{equation}
T_\pm(x) = \sqrt{T_{0\pm}^2\pm\left(1/2\pm\tilde{x}\right) \left(T_{0-}^2 - T_{0+}^2\right)+ \left(\frac{1}{4} - \frac{x^2}{L^2}\right) \frac{V_{ds}^2
}{\mathcal{L}}},
\end{equation}
which presents a maximum a the channel mid-length for large biases. This latter expression reduces  to previous results for diffusive limit \cite{kozub1995} when symmetric boundary conditions are used ($T_{0+}$=$T_{0-}$=$T_0$). When $\tilde{\lambda} = 10$, the voltage drop at the contacts is comparable to that along the channel and the quasi-Fermi level varies along the channel (top of Fig.~\ref{fig:schemes}b). The forward (backward) temperature $T_+$ ($T_-$) increases monotonously as carriers move away from the source (drain) achieving values that under high biases can be significantly higher than the contact temperature $T_{0+}$ ($T_{0-}$) (middle of Fig.~\ref{fig:schemes}b). However intra-branch coupling is not strong enough to restore thermal equilibrium between forward and backward populations. In this case, the quasi-Fermi level separation is still significant compared to the branch thermal broadening, so the local carrier distribution function exhibits a two-step shape which, in the presence of backscattering, are more rounded than in the ballistic limit (bottom of Fig.~\ref{fig:schemes}b).

\begin{figure}[htpb]
\hspace{1in}
\includegraphics[width=4.25in]{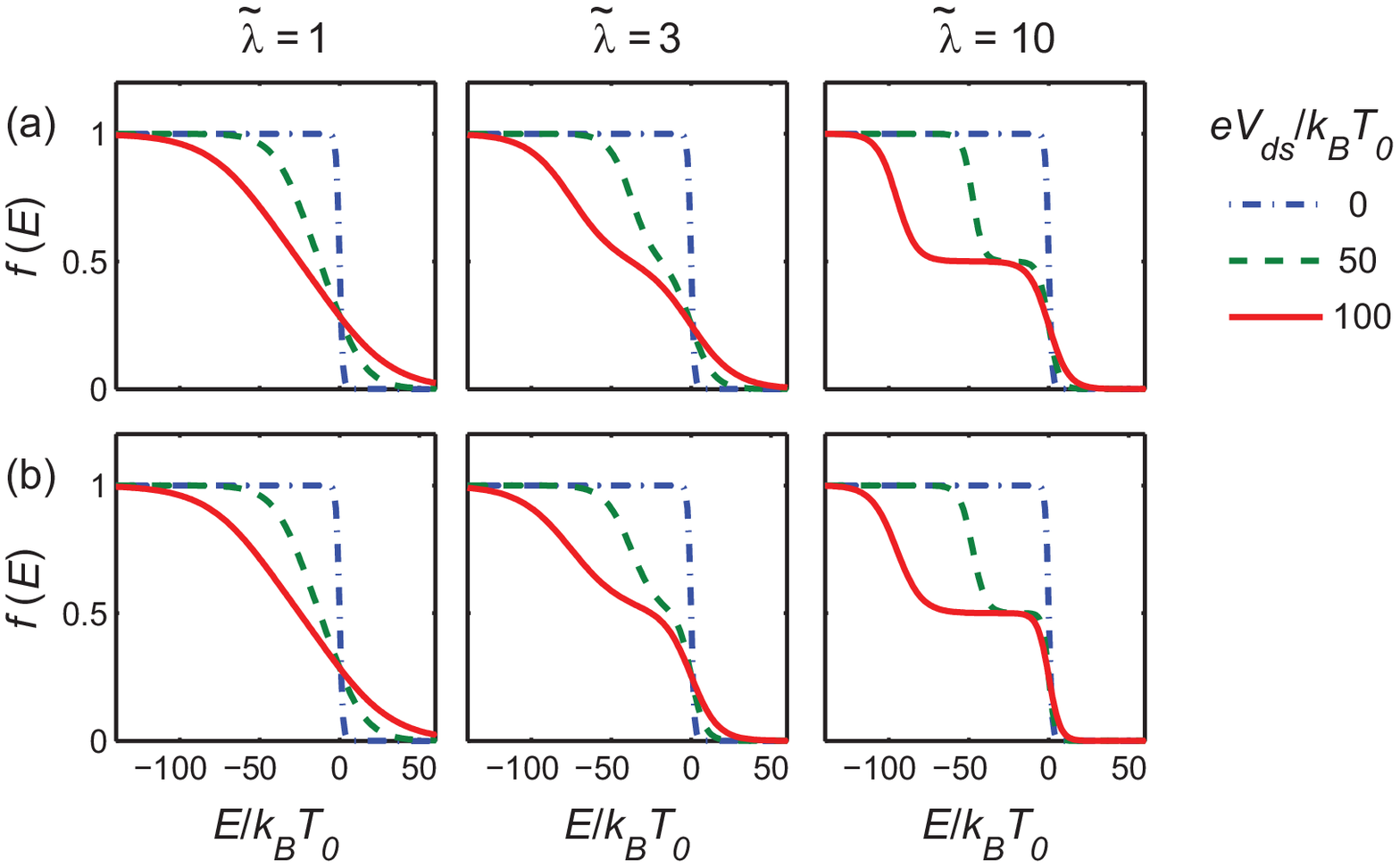}
\caption{Carrier distribution function at the wire $x=0$ (top) and
$x=L/4$ (bottom) for $\tilde{\lambda}$ ratios  of 1, 3 and 10 and bias voltages
$eV_{ds}/k_BT_0$=0, 50 and 100. \label{fig:inter_regime}}
\end{figure}

As mentioned above, systems in which the mean free path becomes comparable to the length are of particular interest because they exhibit features of both the ballistic (splitting of quasi-Fermi levels) and diffusive regime (thermal broadening). In Fig.~\ref{fig:inter_regime}, we show the quasi-particle distribution for the ratios $\tilde{\lambda} = 1$, 3 and 10 at the position $\tilde{x} = 0$ (left column) and -1/4 (right column). We use the symmetric boundary conditions $T_+(-L/2) =T_-( L/2) = T_0$ and bias voltages $eV_{ds}/k_BT_0$ of 0, 50 and 100. As the ratio $\tilde{\lambda}$ increases, the step in the distribution becomes more pronounced as the thermal broadening for forward and backward distribution decreases. At the wire mid-length ($\tilde{x} = 0$), the temperatures of forward and backward distributions have the same value, regardless of the $\tilde{\lambda}$ value. In contrast, the thermal imbalance between forward and backward populations at $\tilde{x} = -1/4$  becomes more prominent as $\tilde{\lambda}$ increases. At this location calculations show that the temperature of the forward population is smaller than that of the backward due to the proximity of the left lead where the carriers moving to the right are injected in the channel. 

As $-1/2\leq\tilde{x}\leq1/2$,  of the forward (backward) carrier temperature (Eq.\ref{eq:tempprof}) achieves its maximum value  at $\tilde{x}=1/2$ ($\tilde{x}=-1/2$) if $\tilde{\lambda}>1$ or $\tilde{x} = \tilde{\lambda}/2$  ($\tilde{x} = -\tilde{\lambda}/2$) if $\tilde{\lambda}\leq1$. For $T_{0+} = T_{0-}=T_0$, the maximum carrier temperature is:
\begin{equation}
 T_{max} = \left\{ \begin{array}{ll}
\sqrt{T_0^2+\frac{V_{ds}^2\tilde{\lambda}}{\mathcal{L}(1+\tilde{\lambda} )^2}}& \textrm{if $\tilde{\lambda} > 1$}\\
\sqrt{T_0^2+\frac{V_{ds}^2}{4\mathcal{L}} }& \textrm{if $\tilde{\lambda} \leq 1$}
 \end{array} \right. .
\end{equation}
Hence, $T_{max}\gg T_0$ for large enough $V_{ds}$, even when $\tilde{\lambda} >1$. In particular, when $eV_{ds}\gg k_BT_0$ the carrier temperature profile (Eq.\ref{eq:tempprof}) becomes:
\begin{equation}
T_\pm(x) = \frac{|V_{ds}|}{1+\tilde{\lambda}} \sqrt{\frac{1}{\mathcal{L}}\left(1/2\pm
\tilde{x}\right)\left(1/2\mp \tilde{x}+\tilde{\lambda}\right)}.\label{eq:scaling_law1}
\end{equation}
Consequently, as both the branch carrier thermal broadening (Eq.~\ref{eq:scaling_law1}) and quasi-Fermi level separation (Eq.~\ref{eq:Vds}) scale linearly with the applied bias  for $eV_{ds} \gg k_BT_0$, the local distribution function (Eq.~\ref{eq:distfunc}) can be expressed in terms of the reduced parameter $E/eV_{ds}$ as experimentally reported by Pothier et al.~\cite{pothier1997}.

We observe that the measurements of the electrical characteristics and the carrier distribution function complement each other in the determination the carrier mean free path in quantum wires. On the one hand, if the system is diffusive ($\tilde{\lambda}\ll 1$), the local thermal imbalance between forward and backward population vanishes (Eq.~\ref{eq:therm_imb}), but the mean free path can be estimated by using Eq.~\ref{eq:Vds} given that most of the voltage drop occurs along the channel. On the other hand, when the mean free path becomes comparable to the length of the system, the determination of $\lambda$ from Eq.~\ref{eq:Vds}, may not be straight forward due to the presence of spurious contact resistances. Nevertheless, the mean free path can be obtained from the forward and backward thermal broadening (Eq.~\ref{eq:tempprof}). Indeed, combining Eq.~\ref{eq:tempprof} for both populations with the symmetric boundary conditions ($T_{0+}=T_{0-}$), we obtain:
\begin{equation}
\frac{\tilde{\lambda}}{(1+\tilde{\lambda})^2} = \tilde{x} \frac{\mathcal{L} \left[T_+(x)^2-T_-(x)^2\right]}{2 V_{ds}^2}, \label{eq:therm_imb}
\end{equation}
where the $T_\pm(x)$ values can be easily extracted from the two step distribution function in the high bias regime ($eV_{ds} \gg k_BT_0$), as shown in Fig.~\ref{fig:inter_regime}.

As the carrier mean free path depends on both the quality of the quantum wires and the number of branches (subbands) involved in carrier transport, a reduction in the number of branches crossing the Fermi level causes a decrease in carrier scattering (and a mean free path increase). Therefore,  metallic wires with larger aspect ratio are well suited for the observation of quasi-ballistic regime features  ($\tilde{\lambda} \gtrsim 1$). However, the larger the aspect ratio, the more significant the effects of surface roughness as a source of scattering. In this regard, pristine carbon nanotubes and graphene nanoribbons are ideal candidates to confirm Eq.~\ref{eq:therm_imb} due to their perfect surfaces and weak interaction with substrates. The extremely high Fermi velocity of carbon nanotubes \cite{saito1992} favors a large mean free path. Unfortunately it simultaneously discretizes the 1D states along the nanotube so that typical micrometer-long nanotubes exhibit 3D confinement as in quantum-dots \cite{bockrath1999, chen2009}. Therefore  centimeter long nanotubes are necessary to achieve a single degree of freedom as in quantum wires. These phenomena can also be observed in graphene nanoribbons with typical widths of a few tens of nanometers \cite{han2007} in field-effect-transistor like configuration. In this case, the Fermi level can be tuned by the back gate voltage to populate successive subconducting bands and change the scattering rate. Finally let us point out that in addition to a temperature lower bound imposed by the level spacing, there is an upper bound that limits measurements due to the critical temperature of the superconducting probe. Moreover, high temperatures not only reduce the mean free path of the system, but also favors the occurrence of dissipative collisions, which is beyond the scope of this work.

In conclusion, we have described a model for the non-coherent hot carrier transport in mesoscopic quantum wires in a situation intermediate between the ballistic and diffusive regimes. The model is in good agreement with experimental findings in mesoscopic copper wires; at the sane time, it provides a reinterpretation of the inhomogeneous thermal broadening in the local carrier distribution function and its scaling with external bias. It offers a direct way to extract the inter-branch carrier mean free path from the temperature imbalance in the non-equilibrium quasi-particle distribution functions. The conditions for the observation of this phenomena in 1D conductors such as nanotubes and graphene nanostructures, are also discussed.

\ack
Marcelo Kuroda acknowledges the support of the Department of Physics at University of Illinois at Urbana-Champaign. We thank Nadya Mason and Yung-Fu Chen for useful discussion.

\section*{References}
\bibliography{report}

\end{document}